\documentclass[aps,prb,twocolumn,groupedaddress,showpacs]{revtex4}
\usepackage{bm}
\usepackage{amssymb}
\usepackage{graphicx,epsfig}
\usepackage{subfigure}

\begin{document}
\title{Controllable Coupling in Phase-Coupled Flux Qubits}
\author{Mun Dae Kim}
\affiliation{Korea Institute for Advanced Study, Seoul 130-722, Korea}
%\date{\today}
\begin{abstract}
%Realistic quantum computing requires two-qubit coupling which is tunable as well as sufficiently strong
%for two-qubit gate operations.
We propose a scheme for tunable coupling  of phase-coupled flux qubits.
The phase-coupling scheme can provide a strong coupling strength
of the order of Josephson coupling energy of Josephson junctions in the connecting loop,
while the previously studied inductive coupling scheme cannot provide
due to small mutual inductance and induced currents.
We show that, in order to control the coupling,
we need {\it two} dc-SQUID's in the connecting loop and
the control fluxes threading the dc-SQUID's  must be in {\it opposite} directions.
The coupling strength is analytically calculated as a function of the control flux at the co-resonance point.
%We have found that  we must introduce two dc-SQUID's with control fluxes in {\it opposite} directions
%in order to control the coupling strength of the phase-coupled qubits which
%can be calculated analytically at the co-resonance point.
\end{abstract}

%\draft
\maketitle

\section{Introduction}

Superconducting Josephson junction qubit is one of the most promising
candidates for implementing quantum computation \cite{Galindo}.
Single qubit coherent oscillations in superconducting  qubits have been demonstrated
experimentally \cite{Nakamura, Yu, Chio} and furthermore two qubit coupling and
entanglement have been performed in charge \cite{Pashkin, Yamamoto}, flux \cite{Izmalkov, Majer, Grajcar}
and phase \cite{Berkley} qubits.  Scalable quantum computing requires controllable and selective coupling
between two remote as well as nearest neighbor qubits.
%Capacitive coupling in charge \cite{Pashkin, Yamamoto} and phase \cite{Berkley} qubits
%can be strong,  but  it is hard to control the coupling strength.
%On the contrary the inductive coupling in flux qubits \cite{Izmalkov, Majer,You,Bertet,Liu} can be tunable,
%but it is too weak to perform efficient two-qubit gate operations.
Recently much theoretical efforts have been devoted on the study about the controllable coupling
of charge \cite{Averin}, charge-phase \cite{Blais} and flux qubits \cite{Niskanen,Liu,Bertet,Plourde}.
For flux qubits  the controllable coupling schemes use inductive coupling,
but it is too weak to perform efficient two-qubit gate operations.
Hence, while in superconducting charge qubit two-qubit coherent oscillations and CNOT gate operations were
experimentally observed \cite{Pashkin, Yamamoto}, only spectroscopy measurement was done for inductively coupled flux qubit \cite{Majer}.
In this study thus we suggest a scheme to give both strong and tunable coupling between two phase-coupled
flux qubits.
The phase-coupling scheme, which we previously proposed \cite{Kim},
has been realized in a recent experiment \cite{Ploeg}.
The controllable coupling scheme using phase-coupled qubits
with threading AC magnetic field was also studied
theoretically \cite{Grajcar2}.
Further, there have been studies about somewhat different phase-coupling schemes
\cite{Grajcar3, Brink}.

Two current states of a flux qubit are characterized by  the induced loop current
related with the phase differences across Josephson junctions in the qubit loop.
If we try to couple two flux qubits using mutual inductance,
the coupling strength $J=MI_LI_R\approx 0.5{\rm GHz}$ \cite{Majer}
will be too weak to perform the discriminating CNOT gate operations \cite{Kim},
since the mutual inductance $M$ and the induced
currents of the left (right) qubit $I_{L(R)}$ is very small.
%induced flux $\Phi_{\rm ind}$ is very weak.
Even though the induced currents of flux qubits are weak, the phase differences
across Josephson junctions $\phi$ are as large as $\phi/2\pi \gtrsim 0.16$.
Hence, if two flux qubits are coupled by the phase differences between two Josephson junctions of different qubits,
we can achieve a  strong coupling of the order of Josephson coupling energy  $E'_J$ of the Josephson
junctions in the connecting loop whose typical value is as large as up to about 200GHz.
%
%since the coupling strength of the phase-coupled flux qubits can be approximately
%written as $4E'_J(1-\cos\phi)$ as will be shown in the following,

Introducing two dc-SQUID's interrupting the connecting loop as shown in Fig. \ref{Coup3JJs}
we can  control the coupling between phase-coupled flux qubits.
The control fluxes, $f'_L$ and $f'_R$, threading two dc-SQUID loops must be in {\it opposite} directions
in order to give rise to the controllable coupling. % as well as switching of the coupled qubits.
When two fluxes are in the {\it same} direction,
the change of control fluxes induces
an additional current flowing in the connecting loop,
causing the shift of qubit states as well as the change of coupling strength.
Such a dilemma also persists in the case of  {\it one} dc-SQUID loop in connecting loop.
However, if the control fluxes are in opposite directions, we have found that
the additional currents coming from two dc-SQUID's are cancelled each other and
thus the coupling strength can be tunable remaining the qubit states unchanged.

\section{Phase-coupling of flux qubits}

\begin{figure}[b]
\vspace{7cm} \includegraphics{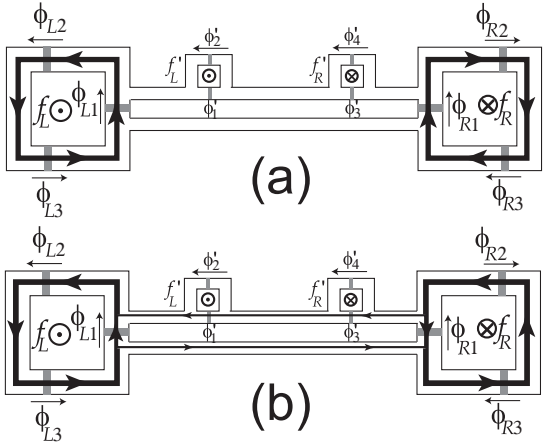}
\vspace*{-1.3cm}
\caption{Phase-coupled flux qubits with a connecting loop interrupted by two dc-SQUID's.
The arrows indicate the flow of the Cooper pairs and thus in reverse direction is the current.
Gray squares denote Josephson junctions with Josephson coupling energy $E_{Ji}$  for qubit loops and
$E'_J$ for connecting loop.
The qubit operating point is $f_L\approx 0.5$ and  $f_R\approx 0.5$.
(a) Two phase-coupled qubits are in the same current state, $|\downarrow\downarrow\rangle$,
where two phases $\phi_{L1}$ and $\phi_{R1}$
are nearly equal to each other resulting negligible coupling energy.
Here $|\downarrow\rangle$ and $|\uparrow\rangle$ denote
the diamagnetic and paramagnetic current state, respectively.
(b) Left (right) qubit is in  the diamagnetic (paramagnetic) current state,
$|\downarrow\uparrow\rangle$, where
the large phase difference, $\phi_{L1}-\phi_{R1}$, induces large Josephson energy
and  current in the connecting loop.
}
\label{Coup3JJs}
\end{figure}

The three-Josephson junctions qubits \cite{Mooij, Orlando, Kim1} in
Fig. \ref{Coup3JJs} has two current states; if the qubit current
$I\propto -E_{Ji}\sin\phi_i <0$, it is diamagnetic while, if
$I>0$, paramagnetic. Introducing the  notation $|\downarrow\rangle$
$(|\uparrow\rangle)$ for diamagnetic (paramagnetic) current state
of a qubit in pseudo spin language, there can be four current
states of coupled qubits, $|\downarrow\downarrow\rangle$,
$|\uparrow\uparrow\rangle$, $|\downarrow\uparrow\rangle$ and
$|\uparrow\downarrow\rangle$, of which we show one of  the same current
states, $|\downarrow\downarrow\rangle$,
and one of the different current  states, $|\downarrow\uparrow\rangle$, in Fig.
\ref{Coup3JJs}.
The phase $\phi_{L1}$ and $\phi_{R1}$ of the Josephson
junctions of the three-Josephson junctions qubits
have different values if two qubits are in different states. Then the phase difference
$\phi_{L1}-\phi_{R1}$ induces the phases  $\phi'_i$ in the Josephson junctions of dc-SQUID loops.

%%%%%%%%%%%%%%%%%%%%%%%%%%%%%%%%%%%%%%%%%%%%%%%%%%%%%%%%%%%%%%%%%%%%%%%%%%%%%%%%%%%%%%%%%%%
%Fig.1
%%%%%%%%%%%%%%%%%%%%%%%%%%%%%%%%%%%%%%%%%%%%%%%%%%%%%%%%%%%%%%%%%%%%%%%%%%%%%%%%%%%%%%%%%%%

%
If we neglect small kinetic inductance, the boundary conditions of the left (right) qubit
and the connecting loop can be approximately written as
\begin{eqnarray}
\label{qubitbc}
\phi_{L(R)1}+\phi_{L(R)2}+\phi_{L(R)3}=2\pi (n_{L(R)}+ f_{t,L(R)}), ~~~~~~~\\
\label{loopbc}
\phi'_1+\phi'_3=2\pi (r+f'_{\rm ind})+(\phi_{L1}-\phi_{R1}),~~~~~~~~~~~~~~~~~~~~\\
\label{switchbc}
-\phi'_1+\phi'_2=2\pi(f'_L+p), ~-\phi'_3+\phi'_4=-2\pi(f'_R+q),~~
\end{eqnarray}
where $f_{t,L(R)}=f_{L(R)}+f_{{\rm ind},L(R)}$ is total flux and
$f_{L(R)}\equiv \Phi_{{\rm ext},L(R)}/\Phi_0$ with the external flux $\Phi_{{\rm ext},L(R)}$ and
the unit flux quantum $\Phi_0=h/2e$ is dimensionless reduced flux threading the left (right) qubit.
Here $f_{{\rm ind},L(R)} \equiv L_sI_{L(R)}/\Phi_0$ with the self inductance $L_s$
and the induced current $I_{L(R)}$ of qubit loop is the induced flux
of each qubit and $f'_{\rm ind} \equiv L'_sI'/\Phi_0$ that of the connecting loop
and $n_L, n_R, r, p$ and $q$ are integers.

We consider that the external fluxes $f_L$ and $f_R$ threading the qubit loops are also
in opposite directions, since they are connected in a twisted way in the scalable design
of Ref. \onlinecite{Kim}. However, for just two qubit coupling, we can choose
the directions of external fluxes threading the qubit loops arbitrary.
Actually there is no external flux in the connecting loop,
but the phase difference $(\phi_{L1}-\phi_{R1})$ in the boundary condition of Eq. (\ref{loopbc})
plays the role of effective flux in the connecting loop, %$f'_{\rm eff}\equiv (\phi_{L1}-\phi_{R1})/2\pi$.
\begin{eqnarray}
\label{feff}
f'_{\rm eff}\equiv\frac{\phi_{L1}-\phi_{R1}}{2\pi}.
\end{eqnarray}
When two qubits are in different current state, i.e., one is diamagnetic and the other
paramagnetic, the value of $f'_{\rm eff}$ becomes $0.3 \sim 0.7$.
Since the induced flux of flux qubit is so weak as $f_{\rm ind}\approx 0.002$,
large value of $f'_{\rm eff} \gg f_{\rm ind}$ in  the phase-coupled
flux qubits can give a strong coupling compared to the inductive coupling scheme.

The  Hamiltonian  of the coupled qubits can be given by
\begin{eqnarray}
\label{Hamiltonian}
\hat{H}=\frac12\hat{{\bf P}}^T\cdot {\bf M}^{-1} \cdot \hat{{\bf P}}
+U_{\rm eff}(\hat{\bm{\phi}}),
\end{eqnarray}
which describes dynamics of a particle with effective mass
$M$ in the effective potential $U_{\rm eff}(\hat{\bm{\phi}})$
with $\bm{\phi}=(\phi_{L1}, \phi_{L2}, \phi_{L3}, \phi_{R1}, \phi_{R2}, \phi_{R3},
\phi'_1, \phi'_2, \phi'_3, \phi'_4)$. %\cite{Orlando}
The kinetic part of the Hamiltonian  comes from the charging energy  of the Josephson junctions
such as
\begin{eqnarray}
E_C(\bm{\phi})=\frac12 \left(\frac{\Phi_0}{2\pi}\right)^2\sum_{P=L,R}
\left(\sum^3_{i=1}C_{Pi}\dot{\phi}^2_{Pi}+C'_P\dot{\phi'}_P^2 \right),
\end{eqnarray}
%$E_C=(1/2)(\Phi_0/2\pi)^2[\sum^3_{i=1}\sum_{P=L,R}C_{Pi}\dot{\phi}^2_{Pi}+C'\dot{\phi'}^2]$,
where $C_{L(R)i}$ and $C'$ is the capacitance of the Josephson junctions of the left (right) qubit loop and
the connecting loop, respectively.
The number of excess Cooper pair charges on Josephson junction $\hat{N_i}$
$\equiv \hat{Q}_i/q_c$
is conjugate to the phase difference $\hat{\phi}_i$ such as $[\hat{\phi}_i,\hat{N_i}]=i$,
where $Q_i=C_i(\Phi_0/2\pi){\dot\phi_i}$, $q_c=2e$ and
 $C_i$ the capacitance of the Josephson junctions.
Here we  introduce the canonical momentum $\hat{P}_i$
%$\hat{P}_i= -i\hbar\partial /\partial \hat{\phi}_i$
and the effective mass $M_{ij}$
%$M_{ij}=(\Phi_0/2\pi)^2C_i\delta_{ij}$
\begin{eqnarray}
\hat{P}_i\equiv \hat{N_i}\hbar= -i\hbar \frac{\partial}{\partial \hat{\phi}_i},~~
M_{ij}=\left(\frac{\Phi_0}{2\pi}\right)^2C_i\delta_{ij},
\end{eqnarray}
to obtain the kinetic part of the Hamiltonian.

The effective potential of the coupled qubits  is composed of
the inductive energy of loops and the Josephson junction energy terms;
%$U_{\rm JJ}(\bm{\phi})=U_{\rm qubit}(\bm{\phi})+U_{\rm conn}(\bm{\phi})$
\begin{eqnarray}
\label{Usum}
U_{\rm eff}(\bm{\phi})&=&U_{\rm ind}(\bm{\phi})+U_{\rm qubit}(\bm{\phi})+U_{\rm conn}(\bm{\phi}),~~~\\
U_{\rm ind}(\bm{\phi})\!\!\!&=&\!\!\!\frac12 L_s(I^2_{L}+I^2_{R})+\frac12 L'I'^2\\
U_{\rm qubit}(\bm{\phi})\!\!\!&=&\!\!\!\sum^3_{i=1}E_{Ji}(1-\cos\phi_{Li})\!\!+\!\!\sum^3_{i=1}E_{Ji}(1-\cos\phi_{Ri})\nonumber\\
\\
U_{\rm conn}(\bm{\phi})\!\!\!&=&\!\!\!\sum^4_{i=1}E'_{J}(1-\cos\phi'_i).
\end{eqnarray}
Here $U_{\rm ind}(\bm{\phi})$ is the inductive energy of loops with
the current of the right qubit $I_R$, left qubit $I_L$ and connecting loop $I'$.
$U_{\rm qubit}(\bm{\phi})$
is the energy of the Josephson junctions in two qubit loop
and $U_{\rm conn}(\bm{\phi})$ that of the connecting loop
with Josephson coupling energies $E_{Ji}$ and $E'_J$.

In experiments the two Josephson junctions with phase differences $\phi_{L(R)2}$ and $\phi_{L(R)3}$
can be considered nominally the same  so that it is reasonable to set
%$E_{J2}=E_{J3}=E_J$ and $\phi_{L(R)2}=\phi_{L(R)3}.$
\begin{eqnarray}
\label{J23}
E_{J2}=E_{J3}=E_J,~~~\phi_{L(R)2}=\phi_{L(R)3}.
\end{eqnarray}
Here we introduce a rotated coordinate
\begin{eqnarray}
\phi_p = (\phi_{L3}+\phi_{R3})/2 ,\\
\phi_m = (\phi_{L3}-\phi_{R3})/2
\end{eqnarray}
and then using the boundary conditions in Eq. (\ref{qubitbc})
we get
\begin{eqnarray}
\label{phi1}
\phi_{L1}\pm \phi_{R1}=-4\phi_{p(m)}+2\pi(n_L\pm n_R+f_L \pm f_R).
\end{eqnarray}
Thus we can reexpress the sum of Josephson junction energies of both qubits
as
%$U_{\rm qubit}(\bm{\phi})=2E_{J1}[1-\cos(P\pi-2\phi_p)\cos(M\pi-2\phi_m)]+4E_J(1-\cos\phi_p\cos\phi_m)],$
\begin{eqnarray}
\label{Ecoup}
U_{\rm qubit}(\bm{\phi})=2E_{J1}[1-\cos(P\pi-2\phi_p)\cos(M\pi-2\phi_m)]\nonumber\\
+4E_J(1-\cos\phi_p\cos\phi_m),~~~~~~~~~~~~~~~~~~~~~~~~~~~~
\end{eqnarray}
where $P\equiv n_L+n_R+f_L+f_R+f_{{\rm ind},L}+f_{{\rm ind},R}$ and
$M\equiv n_L-n_R+f_L-f_R+f_{{\rm ind},L}-f_{{\rm ind},R}$.
Since experimentally qubit operations are performed at near the co-resonance point
$f_L=f_R=0.5$ and the induced flux is so weak as $f_{\rm ind, L(R)}\approx 0.002$,
$P$ and $M$ can be approximated as
integers such that $P= n_L+n_R+1$ and $M=n_L-n_R$. If $P$ is even, $M$ is odd and vise versa,
so we can get simple form for $U_{\rm qubit}(\bm{\phi})$,
\begin{eqnarray}
\label{Uqubit}
U_{\rm qubit}(\phi_m,\phi_p)=2E_{J1}\cos2\phi_p\cos2\phi_m ~~~~~~~~~~~~~~~~\nonumber\\
-4E_J\cos\phi_p\cos\phi_m  +2E_{J1}+4E_J. ~~~~~~~~
\end{eqnarray}

Introducing another rotated coordinate
\begin{eqnarray}
\phi'_p = (\phi'_{1}+\phi'_{3})/2 ,\\
\phi'_m = (\phi'_{1}-\phi'_{3})/2
\end{eqnarray}
 and using the boundary conditions in Eq. (\ref{switchbc})
to get
\begin{eqnarray}
\label{phi'2}
(\phi'_2\pm \phi'_4)/2=\phi'_{p(m)}+\pi(f'_L\mp f'_R+p\mp q),
\end{eqnarray}
the Josephson junction energy of the connecting loop $U_{\rm conn}(\bm{\phi})=\sum^4_{i=1}E'_{J}(1-\cos\phi'_i)$
can also be written as
\begin{eqnarray}
\label{Uswi}
U_{\rm conn}(\phi'_m,\phi'_p)/2E'_J=2-\cos\phi'_m\cos\phi'_p~~~~~~~~~~~~~~ \\
-\cos[\phi'_m+\pi(f'_L+f'_R)]\cos[\phi'_p+\pi(f'_L-f'_R)],\nonumber
\end{eqnarray}
where we set $p=0$ and $q=0$.

%\begin{eqnarray}
%&&E'_J(1-\cos\phi'_1)+E'_J(1-\cos\phi'_2)=2E'_J(1-\cos((\phi'_1-\phi'_2)/2)\cos((\phi'_1+\phi'_2)/2))\\
%&+&2E'_J[1-\cos\phi'_m\cos(M'\pi-2\phi_m)\\
%&+&E'_J(1-Q\cos2\phi_m)
%\end{eqnarray}

\section{Coupled-qubit states in effective potential}

First of all we  consider the case that the control fluxes, $f'_L$ and $f'_R$, have opposite
directions such that
\begin{equation}
f'_L=f'_R=f'.
\end{equation}
Note that the boundary conditions in  Eq. (\ref{switchbc}) already have opposite signs.
In order to  obtain the effective potential
as a function of $\phi_m$ and $\phi_p$, we reexpress $\phi'_p$
as $\phi'_p=\pi(f'_{\rm eff}+r+f'_{\rm ind})=-2\phi_m+\pi M'$
using the boundary conditions in Eq. (\ref{loopbc})  and the expression in Eq. (\ref{phi1}).
Here  $M'\equiv M+r+f'_{\rm ind}$ can be
written as $M'=n_L-n_R+r$ neglecting small induced flux $f'_{\rm ind}$ in the connecting loop.
Depending on whether $M'$ is even or odd, the results will  be quantitatively different, but
qualitatively the same. Here and after, thus, we choose $M'$ is even and specifically $n_L=0$, $n_R=0$
and $r=0$ for simplicity and then $\phi'_p$ becomes
\begin{eqnarray}
\label{rel}
\phi'_p=\pi f'_{\rm eff}=-2\phi_m,
\end{eqnarray}
and
the energy of Josephson junctions in connecting loop in Eq. (\ref{Uswi})
\begin{eqnarray}
\label{Uswidiff}
U_{\rm conn}(\phi'_m,\phi_m)~~~~~~~~~~~~~~~~~~~~~~~~~~~~~~~~~~~~~~\nonumber\\
=4E'_J[1-\cos\pi f'\cos(\phi'_m+\pi f')\cos2\phi_m].
\end{eqnarray}

%%%%%%%%%%%%%%%%%%%%%%%%%%%%%%%%%%%%%%%%%%%%%%%%%%%%%%%%%%%%%%%%%%%%%%%%%%%%%%%%%%%%%%%%%%%
%Fig. 2
\begin{figure}[t]
\vspace{4.5cm}
\includegraphics{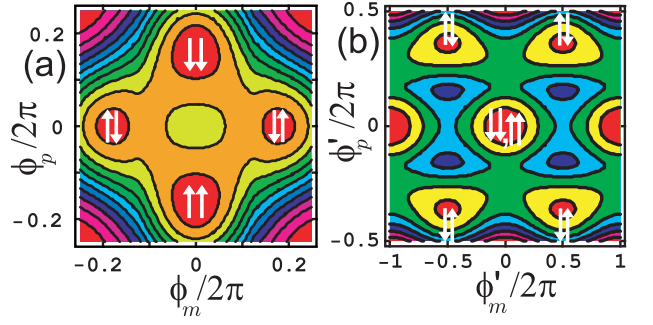}
\hspace{0cm}
%\special{psfile=PotPx.eps vscale=45 hscale=45 hoffset=-25 voffset=2 angle=0}
\vspace{-0.cm}
\caption{(Color online) Effective potential of the coupled qubits in Eq. (\ref{Ump}) for $E'_J=0.1E_J$
when $f'_L=f'_R=f'=0$  (a) in $(\phi_m,\phi_p)$ plane and (b) in $(\phi'_m,\phi'_p)$ plane.
Coupled qubit states at the local minima of potentials are denoted in pseudo-spin notation, which shows that these states are
stable in both planes. Here  we set $E_{J1}=E_J$ and $f_L=f_R=0.5$. }
\label{Potential}
\end{figure}
%%%%%%%%%%%%%%%%%%%%%%%%%%%%%%%%%%%%%%%%%%%%%%%%%%%%%%%%%%%%%%%%%%%%%%%%%%%%%%%%%%%%%%%%%%%

Since the induced energy $U_{\rm ind}(\bm{\phi})$ can be negligible,
the total effective potential $U_{\rm eff}(\bm{\phi})$
in Eq. (\ref{Usum}) is given by the sum of the energies in Eqs. (\ref{Uqubit}) and (\ref{Uswidiff}),
%$U_{\rm eff}(\bm{\phi})=U_{\rm qubit}(\phi_m,\phi_p)+U_{\rm conn}(\phi'_m,\phi_m)$.
\begin{eqnarray}
\label{Ump}
U_{\rm eff}(\phi_m,\phi_p,\phi'_m)\!\!=\!\! U_{\rm qubit}(\phi_m,\phi_p)\!\! +\!\! U_{\rm conn}(\phi'_m,\phi_m).
\end{eqnarray}
The lowest energy level of $U_{\rm eff}$ in $(\phi_m,\phi_p)$ plane can be obtained
by setting the remaining variable $\phi'_m$ in Eq.  (\ref{Uswidiff}) as
\begin{eqnarray}
\label{i}
({\rm i})&&\!\!\! \phi'_m=0, ~~{\rm for}\; -\pi/4 < \phi_m< \pi/4, \\
\label{ii}
({\rm ii})&&\!\!\! \phi'_m=\pm\pi,~~ {\rm for}\; \pi/4<|\phi_m|<\pi/2,
\end{eqnarray}
when $f'=0$.
We plot the effective potential $U_{\rm eff}(\phi_m,\phi_p)$
in Fig. \ref{Potential} (a) with four local minima.

The value of local minima of case (i) can be obtained from $U_{\rm eff}(\phi_m=0,\phi_p)$
and we have found that two local minima  have the same value, $E_{ss}(f')$,
for equal pseudo-spin state with $s\in\{\downarrow, \uparrow\}$.
%of case (i) can be obtained from $U_{\rm eff}(\phi_m=0,\phi_p)$
Similarly we get $E_{s,-s}(f')$ of case (ii)  from $U_{\rm eff}(\phi_m,\phi_p=0)$
for different pseudo-spin state.
As a result, we obtain %$U_{\rm min, 1}=E_J(4-E_J/E_{J1})+4E'_J(1-\cos\pi f')$
\begin{eqnarray}
\label{Umin}
E_{ss'}(f')&=&4E_J+4E'_J(1-\cos\pi f')\nonumber\\
&-&2E_J\cos\phi_{ss'},
\end{eqnarray}
where $\phi_{ss}$ is the value of $\phi_{p}$ at local minima
of the same spin states, $|ss\rangle$,
and $\phi_{s,-s}$ the value of $\phi_{m}$
of the different spin states, $|s,-s\rangle$ with
\begin{eqnarray}
\cos\phi_{ss}&=&\frac{E_J}{2E_{J1}}, \\
\cos\phi_{s,-s}&=&\frac{E_J}{2(E_{J1}+2E'_{J}\cos\pi f')}.
\end{eqnarray}
Thus the energy of the same spin states, $E_{ss}(f'=0)$, is lower than
that of different spin states, $E_{s,-s}(f'=0)$,
as shown in Fig. \ref{Potential}(a).

Here  we set $E_{J1}=E_J$, $E'_J=0.1E_J$ and $f_L=f_R=0.5$.
For the same spin states
we have two solutions, $\phi_{ss}/2\pi= \pm 1/6$, corresponding to two local minima,
$E_{\downarrow\downarrow}(f'=0)$ and $E_{\uparrow\uparrow}(f'=0)$.
When $\phi_{ss}/2\pi=1/6$, $\phi_p/2\pi=1/6$ and $\phi_m=0$ and thus
\begin{eqnarray}
\phi_{L(R)2}/2\pi=\phi_{L(R)3}/2\pi=\phi_{L(R)1}/2\pi=1/6
\end{eqnarray}
using $\phi_{L(R)1}+2\phi_{L(R)3}=\pi$
from the boundary condition in Eq. (\ref{qubitbc}) with $n_{L(R)}=0$.
Since the loop currents of both qubits then
\begin{eqnarray}
I=-(2\pi/\Phi_0)E_{J}\sin\phi_{L(R)3},
\end{eqnarray}
are diamagnetic as can be seen from Fig. \ref{Coup3JJs}(a),
this coupled-qubits state can be represented as $|\downarrow\downarrow\rangle$
as shown in Fig. \ref{Potential}(a).
On the other hand, when $\phi_{ss}/2\pi=-1/6$,
\begin{equation}
\phi_{L(R)2}/2\pi=\phi_{L(R)3}/2\pi=-1/6,  ~\phi_{L(R)1}/2\pi=5/6.
\end{equation}
Then the qubit current
$I=-(2\pi/\Phi_0)E_{J}\sin(-\pi/3)=-(2\pi/\Phi_0)E_{J}\sin(5\pi/3)$
corresponds to the paramagnetic current states, $|\uparrow\uparrow\rangle$.
We would like to note that, since the external fluxes $f_L$ and $f_R$ threading left and right qubit loops  are already
in opposite directions, diamagnetic (paramagnetic) currents of both qubits in $|\downarrow\downarrow\rangle$ ($|\uparrow\uparrow\rangle$)
state are also in opposite directions.

%%%%%%%%%%%%%%%%%%%%%%%%%%%%%%%%%%%%%%%%%%%%%%%%%%%%%
\begin{table}[b]
\begin{center}
\begin{tabular}{c|c|c|c}
\hline
\hline
$|ss'\rangle$& ($\phi_{R3}/2\pi,\phi_{L3}/2\pi$) & ($\phi_m/2\pi,\phi_p/2\pi$) & ($\phi'_m/2\pi,\phi'_p/2\pi$)
\\
\hline\hline
$|\downarrow\downarrow\rangle$ &(1/6,1/6) & ($0,1/6$) & (0,0)
\\
$|\uparrow\uparrow\rangle$ & ($-1/6,-1/6$) & ($0,-1/6$) & (0,0)
\\
 $|\downarrow\uparrow\rangle$ & ($-0.181,0.181$) & ($0.181,0$)& ($\pm 0.5,-0.362$)
\\
$|\uparrow\downarrow\rangle$ & ($0.181,-0.181$) & ($-0.181,0$) & ($\pm 0.5,0.362$)
\\
\hline \hline
\end{tabular}
\end{center}
\caption{The values of phase differences of coupled qubits states in several coordinates with
$E_{J1}=E_J$, $E'_J=0.1E_J$, $f_L=f_R=0.5$ and $f'=0$.}
\label{table}
\end{table}
%%%%%%%%%%%%%%%%%%%%%TABLE%%%%%%%%%%%%%%%%%%%%%%%%%%

%%%%%%%%%%%%%%%%%%%%%%%%%%%%%%%%%%%%%%%%%%%%%%%%%%%%%
%\begin{table}[b]
%\begin{center}
%\begin{tabular}{c|c|c|c}
%\hline
%\hline
%$|ss'\rangle$& ($\phi_{L1}/2\pi,\phi_{R1}/2\pi$) & ($\phi_p/2\pi,\phi_m/2\pi$) & ($\phi'_p/2\pi,\phi'_m/2\pi$)
%\\
%\hline\hline
%$|\downarrow\downarrow\rangle$ &(1/6,1/6) & ($1/6,0$) & (0,0)
%\\
%$|\uparrow\uparrow\rangle$ & (5/6,5/6) & ($-1/6,0$) & (0,0)
%\\
% $|\downarrow\uparrow\rangle$ & (0.138,0.862) & ($0,0.181$)& ($-0.362,\pm 0.5$)
%\\
%$|\uparrow\downarrow\rangle$ & (0.862,0.138) & ($0,-0.181$) & ($0.362,\pm 0.5$)
%\\
%\hline \hline
%\end{tabular}
%\end{center}
%\caption{The values of phase differences of coupled qubits states in several coordinates with
%$E_{J1}=E_J$, $E'_J=0.1E_J$, $f_L=f_R=0.5$ and $f'=0$.}
%\label{table}
%\end{table}
%%%%%%%%%%%%%%%%%%%%%TABLE%%%%%%%%%%%%%%%%%%%%%%%%%%

For different spin states, two solutions are also obtained
for $\phi_{s,-s}/2\pi\approx \pm 0.181$.
When $\phi_{s,-s}/2\pi\approx 0.181$,
\begin{eqnarray}
\phi_{L2}/2\pi=\phi_{L3}/2\pi\approx 0.181,~~~~~
 \phi_{L1}/2\pi\approx 0.138\\
\phi_{R2}/2\pi=\phi_{R3}/2\pi\approx -0.181,~~
\phi_{R1}/2\pi\approx 0.862
\end{eqnarray}
for left and right qubit respectively,
which corresponds to the state, $|\downarrow\uparrow\rangle$
in Fig. \ref{Potential} (a).
In the same way
$\phi_{s,-s}/2\pi\approx -0.181$
corresponds to the state $|\uparrow\downarrow\rangle$.
Hence we can identify four stable states,
$|\downarrow\downarrow\rangle, |\uparrow\uparrow\rangle, |\downarrow\uparrow\rangle$
and $|\uparrow\downarrow\rangle$, with energies
$E_{ss}$ and $E_{s,-s}$ at the local minima of $U_{\rm eff}(\phi_m,\phi_p)$
as shown in Fig. \ref{Potential} (a).

Even though  above four states are stable states in $(\phi_m, \phi_p)$ plane, it can be unstable
in the other dimensions if they are saddle points.  Thus we need represent the effective potential
$U_{\rm eff}$ in $(\phi'_m, \phi'_p)$ plane. From the expression of $U_{\rm eff}(\phi_m,\phi_p,\phi'_m)$
in Eq. (\ref{Ump}) and the relation $\phi'_p=-2\phi_m$ in Eq. (\ref{rel}), we can get $U_{\rm eff}(\phi'_m,\phi'_p,\phi_p)$
and, following similar procedure as in the $(\phi_m, \phi_p)$ plane, we obtain the effective potential
as shown in Fig. \ref{Potential} (b), where we can again see local minima.
In Figs. \ref{Potential} (b), for  the states $|\downarrow\downarrow\rangle$ and $|\uparrow\uparrow\rangle$
of case (i), we can get the values
\begin{eqnarray}
\phi'_p=0~~ {\rm and}~~ \phi'_m=0,
\end{eqnarray}
and, for  case (ii),  the values
\begin{eqnarray}
\left\{
\matrix{
\phi'_p/2\pi=-0.362,  & \phi'_m/2\pi=\pm 0.5, & {\rm for}~~  |\downarrow\uparrow\rangle  \nonumber\\
\phi'_p/2\pi=0.362,  & \phi'_m/2\pi=\pm 0.5, & {\rm for}~~  |\uparrow\downarrow\rangle
}
\right.
\end{eqnarray}
using Eqs. (\ref{rel}), (\ref{i}) and (\ref{ii}) and the values of $\phi_m$ in each case. %,from $\phi_m/2\pi=\pm 0.18$.
As a result, we are able to identify the spin states at local minima of Figs. \ref{Potential} (b)
from Fig. \ref{Potential} (a) with above values and confirm the stability of the states in both planes.
In Table \ref{table} we summarize  the values of the phase differences
for four states, $|ss'\rangle$,
of coupled qubits in several coordinates.
Actually we obtained higher energy states in Fig. \ref{Potential} (a), but
found that they are unstable in $(\phi'_m, \phi'_p)$ plane.
%\begin{eqnarray}
%U_{\rm eff}(\phi'_m,\phi'_p,\phi_p)=2E_{J1}\cos2\phi_p\cos\phi'_p-4E_J\cos\phi_p\cos0.5\phi'_p
%-4E'_J\cos\pi f'\cos(\phi'_m+\pi f')\cos\phi'_p
%\end{eqnarray}

\section{Tunable Coupling of Flux Qubits}

The Hamiltonian of coupled qubits  can be written as
\begin{eqnarray}
\label{Hcoup}
H_{\rm coup}&=&h_L\sigma^z_L\otimes I +h_R I\otimes\sigma^z_R -J\sigma^z_L\otimes \sigma^z_R \nonumber\\
&+&t_L\sigma^x_L\otimes I +t_R I\otimes\sigma^x_R+E_0,
\end{eqnarray}
where $h_L\equiv (E_{\uparrow\downarrow}+E_{\uparrow\uparrow})/2-E_0$
and $h_R\equiv (E_{\downarrow\uparrow}+E_{\uparrow\uparrow})/2-E_0$
with $E_0\equiv (E_{\downarrow\uparrow}+E_{\uparrow\downarrow}+E_{\downarrow\downarrow}+E_{\uparrow\uparrow})/4$
and $I$ is the $2\times 2$ identity matrix.
First two terms are qubit terms, the third is coupling term and last two terms
are tunnelling terms which come from the quantum fluctuation
described by the kinetic term of the Hamiltonian.
Then the coupling constant $J$ of the coupled qubits is given by \cite{Kim}
%$J=(1/4)(E_{\downarrow\uparrow}+E_{\uparrow\downarrow}-E_{\downarrow\downarrow}-E_{\uparrow\uparrow})$ \cite{Kim}.
\begin{eqnarray}
\label{J}
J=\frac14(E_{\downarrow\uparrow}+E_{\uparrow\downarrow}-E_{\downarrow\downarrow}-E_{\uparrow\uparrow}).
\end{eqnarray}
%

%%%%%%%%%%%%%%%%%%%%%%%%%%%%%%%%%%%%%%%%%%%%%%%%%%%%%%%%%%%%%%%%%%%%%%%%%%%%%%%%%%%%%%%%%%%
%Fig. 3
\begin{figure}[t]
\vspace{6cm}
\includegraphics{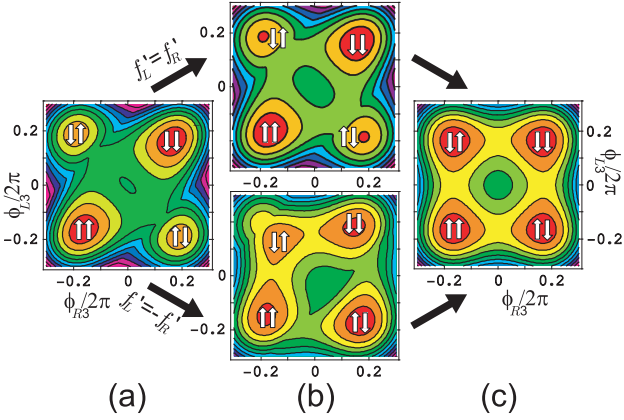}
\vspace{-0.1cm}
\caption{(Color online)
Effective potential of the coupled qubits in Eq. (\ref{Usum}) for $E'_J=0.1E_J$,
$E_{J1}=E_J$ and $f_L=f_R=0.5$. Coupled qubit states at the local minima of potentials
are denoted in pseudo-spin notation.
(a) Effective potential as a function of $\phi_{R3}$ and $\phi_{L3}$ when $f'_L=f'_R=f'=0$
for the phase-coupled qubits in Fig. \ref{Coup3JJs}.
Here the energies of different current states are equal to each other,
$E_{\downarrow\uparrow}=E_{\uparrow\downarrow}$, as well as
$E_{\downarrow\downarrow}=E_{\uparrow\uparrow}$ for the same current states.
The energy of different current states
$E_{s,-s}=E_{\downarrow\uparrow}=E_{\uparrow\downarrow}$ is
higher than that of the same current states
$E_{ss}=E_{\downarrow\downarrow}=E_{\uparrow\uparrow}$.
(b) (top) Two control fluxes in Fig. \ref{Coup3JJs} are in opposite
directions, $f'_L=f'_R=f'$, and $f'$ is increased to $f'=0.25$.
The energy difference $\Delta E=E_{s,-s}-E_{ss}$
becomes smaller than when $f'=0$ in (a).
(bottom) For the case when two control fluxes are in the same direction
such as $f'_L=-f'_R=f'=0.25$,
the energies of different current states are not equal to each other any more;
$E_{\downarrow\uparrow} > E_{\uparrow\downarrow}$.
(c) The coupling becomes switched off when $f'_L=|f'_R|=f'=0.5$.
Thus the energies of four states have the same value,
$E_{\downarrow\downarrow}=E_{\uparrow\uparrow}=
E_{\downarrow\uparrow}=E_{\uparrow\downarrow}$.
}
\label{ContPot}
\end{figure}
%%%%%%%%%%%%%%%%%%%%%%%%%%%%%%%%%%%%%%%%%%%%%%%%%%%%%%%%%%%%%%%%%%%%%%%%%%%%%%%%%%%%%%%%%%%

In Fig. \ref{ContPot} we plot the energies of coupled-qubits
for various $f'$ with $E'_J=0.1E_J$, $E_{J1}=E_J$ and $f_L=f_R=0.5$
in $(\phi_{R3},\phi_{L3})$ plane.
When $f'=0$ in Fig. \ref{ContPot}(a), the energies $E_{ss}$ of the same spin states,
$|\downarrow\downarrow\rangle$ and $|\uparrow\uparrow\rangle$, are lower than
$E_{s,-s}$, of the different spin states,
$|\downarrow\uparrow\rangle$ and $|\uparrow\downarrow\rangle$.
The positions of four local minima are shown in Table \ref{table}.
As increases $f'$, the energy difference $\Delta E=E_{s,-s}-E_{ss}$
becomes smaller (upper panel in (b)) and finally
$\Delta E=0$ at $f'=0.5$ in (c).
Since $E_{\downarrow\downarrow}=E_{\uparrow\uparrow} =E_{ss}$ and
$E_{\downarrow\uparrow}=E_{\uparrow\downarrow} =E_{s,-s}$,
the coupling strength can be written as
\begin{eqnarray}
2J(f') = \Delta E(f')= E_{s,-s}(f')-E_{ss}(f').
\end{eqnarray}
Therefore the coupling strength between two flux qubits changes
as varying the control fluxes $f'$ threading the dc-SQUID loop in the
connecting loop.

\begin{figure}[t]
\vspace{3.8cm}
\includegraphics{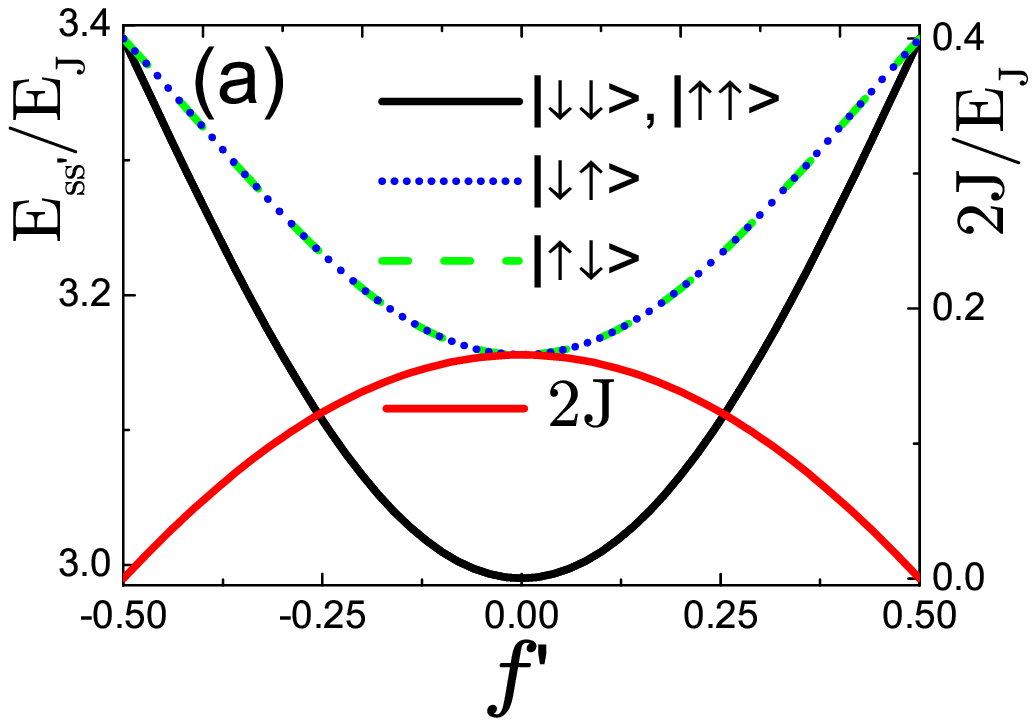}
\vspace{2.7cm}
\includegraphics{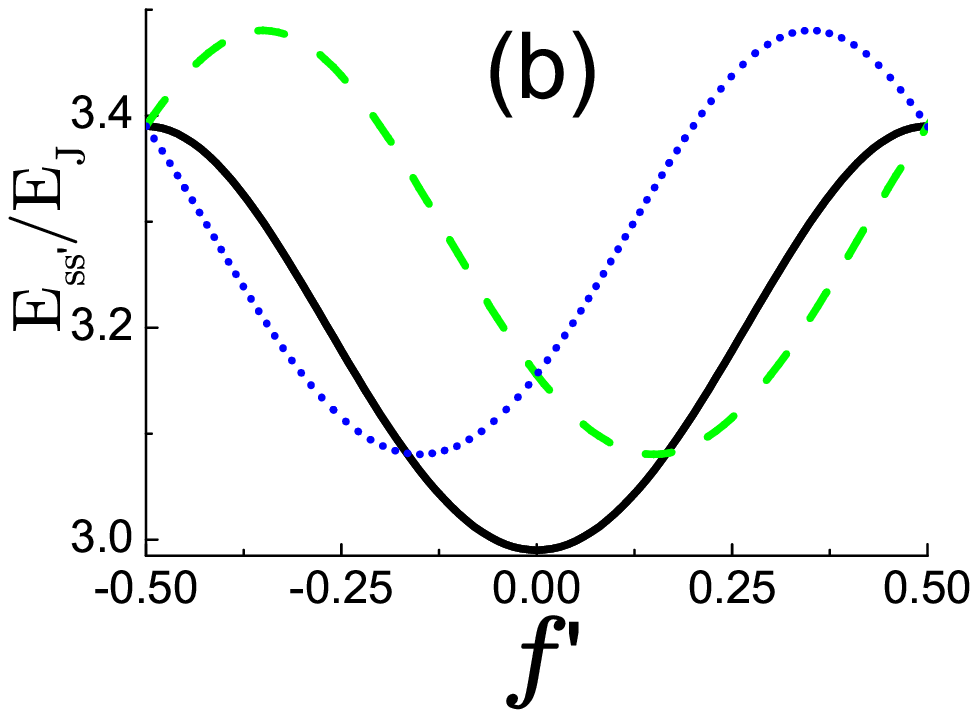}
\includegraphics{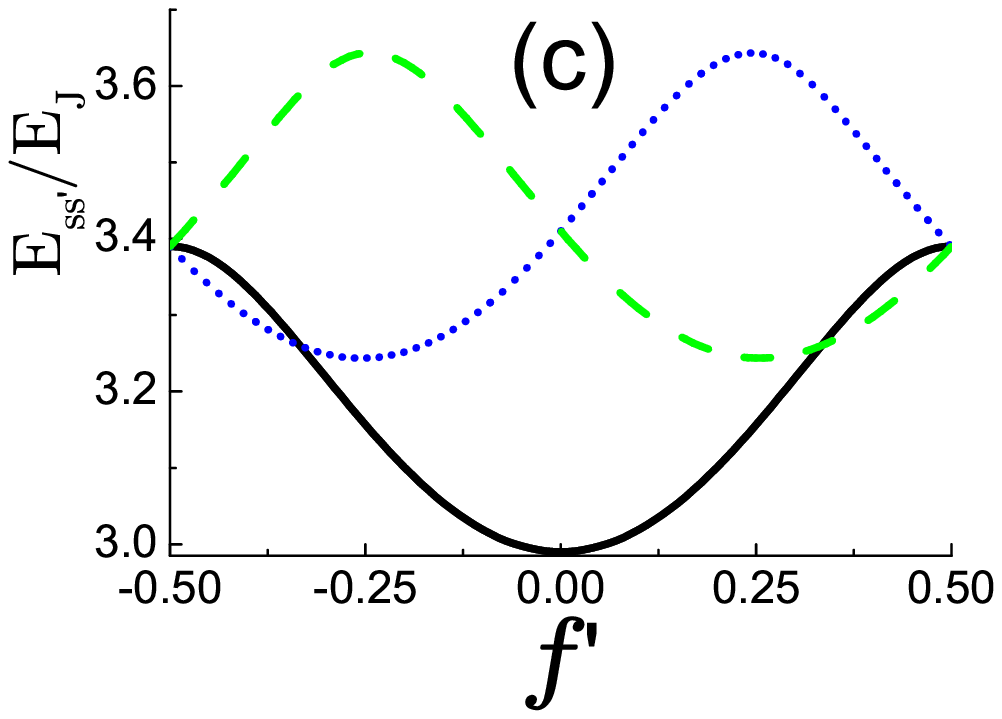}
\vspace{-0.2cm}
\caption{(Color online)
Energies of coupled qubit states for $E'_J=0.1E_J$,
$E_{J1}=E_J$ and $f_L=f_R=0.5$ as a function of $f'$.
(a) $E_{ss'}(f')$ in Eq. (\ref{Umin})
when two control fluxes in Fig. \ref{Coup3JJs} are in opposite directions,  $f'_L=f'_R=f'$.
The coupled qubits can be described by the Hamiltonian in Eq. (\ref{Hcoup})
and the coupling strength
$2J(f') = E_{s,-s}(f')-E_{ss}(f')$
in Eq. (\ref{J}) is also shown.
As increases $f'$, the coupling strength decreases monotonously, vanishing finally at $f'=0.5$.
(b) When two control fluxes in Fig. \ref{Coup3JJs} are in the same direction, $f'_L=-f'_R=f'$,
the energy  of two states, $|\downarrow\uparrow\rangle$ and  $|\uparrow\downarrow\rangle$,
becomes different, as if additional fluxes, $\delta f_L$ and $\delta f_R$,
were applied into the left and right qubit loop, respectively.
Hence the coupling between two qubits cannot be represented  solely by
change of the coupling constant of  the Hamiltonian in Eq. (\ref{Hcoup}).
(c) For the case when there is only one dc-SQUID loop in the connecting loop
instead of two dc-SQUID's in Fig. \ref{Coup3JJs},
the energies of the different current states, $|\downarrow\uparrow\rangle$ and  $|\uparrow\downarrow\rangle$,
are also different from each other.}
\label{TwoDiff}
\vspace{-0.3cm}
\end{figure}

From Eq. (\ref{Umin})  the coupling constant $J$ can be represented as a function of $f'$
by $J(f')=E_J(\cos\phi_{ss}-\cos\phi_{s,-s})$, which gives
\begin{eqnarray}
\label{J}
J(f')=\frac{E^2_J}{E_{J1}}\frac{E'_J\cos\pi f'}{E_{J1}+2E'_J\cos\pi f'}.
\end{eqnarray}
In Fig. \ref{TwoDiff}(a) we plot the energies $E_{ss'}(f')$
and $J(f')$ as a function of $f'$,
where $2J(f') = E_{s,-s}(f')-E_{ss}(f')$.
When $f'=0$, $J$ is of the order  of $E'_J$ so that we can obtain a sufficiently strong coupling.
%by using Josephson junctions with high coupling energy.
By adjusting $f'$ the coupling strength can be tuned
from strong coupling to zero at $f'=0.5$.

The coupling strength $J(f')$ in Eq. (\ref{J}) depends on $E_J/E_{J1}$ as well as  $E'_J$.
When $E'_J/E_{J1}$ is small, $J(f')$ is proportional to $E'_J$ and $(E_J/E_{J1})^2$.
Recently the phase-coupling scheme has been experimentally
implemented \cite{Ploeg}, where  four-Josephson junctions qubits are employed
instead of usual three-Josephson junctions qubits.
In that experiment
the Josephson junction energy $E_{J1}$ of fourth junction is large so that
the value of $E_J/E_{J1}$ is about $E_J/E_{J1} \approx 1/3$.
As a result, the experiment exhibits rather small
coupling strength.

%%%%%%%%%%%%%%%%%%%%%%%%%%%%%%%%%%%%%%%%%%%%%%%%%%%%%%%%%%%%%%%%%%%%%%%%%%%%%%%%%%%%%%%%%%%
%Fig. 4
%%%%%%%%%%%%%%%%%%%%%%%%%%%%%%%%%%%%%%%%%%%%%%%%%%%%%%%%%%%%%%%%%%%%%%%%%%%%%%%%%%%%%%%%%%%

The current of connecting loop can be written as $-(\Phi_0/2\pi)I'=E'_J\sum^2_{i=1}\sin\phi'_i=E'_J\sum^4_{i=3}\sin\phi'_i$,
which gives the relations, $(\phi'_1-\phi'_3)+(\phi'_2-\phi'_4)=4\pi k$ and then
\begin{equation}
\label{k}
\phi'_m=(k-f')\pi,
\end{equation}
with integer $k$
using the boundary conditions in Eq. (\ref{switchbc}).
Then, using Eq. (\ref{phi'2}) and the effective flux $f'_{\rm eff}$  in Eq. (\ref{rel}),
the current $-(\Phi_0/2\pi)I'=0.5E'_J\sum^4_{i=1}\sin\phi'_i$ is given by
\begin{eqnarray}
\label{I'}
I'=-\left(\frac{2\pi}{\Phi_0}\right)(-1)^k 2E'_J\cos\pi f'\sin\pi f'_{\rm eff}.
\end{eqnarray}
This current-phase relation can be considered as the Josephson junction type relation,
$I'=-(2\pi/\Phi_0)(-1)^k \tilde{E}'_J\sin\varphi$,
with the effective Josephson coupling energy, $\tilde{E}'_J$, of two dc-SQUID's in the connecting
loop
\begin{eqnarray}
\tilde{E}'_J= 2E'_J\cos\pi f'
\end{eqnarray}
and the phase difference $\varphi=\pi f'_{\rm eff}$.
The coupling constant in Eq. (\ref{J}) also can be represented by
the effective Josephson coupling energy, $\tilde{E}'_J$.
Thus the large phase difference, $\pi f'_{\rm eff}$, and the Josephson coupling energy, $\tilde{E}'_J$,
induce the current in the connecting loop and  the coupling energy of the phase-coupled qubits.

For the same spin states,  $|\downarrow\downarrow\rangle$ and $|\uparrow\uparrow\rangle$,
the current of connecting loop $I'$ becomes zero, since $\phi_{L1}=\phi_{R1}$ and thus $f'_{\rm eff}=0$.
For a different spin states $|\downarrow\uparrow\rangle$ with $f'=0$,
$f'_{\rm eff}=(\phi_{L1}-\phi_{R1})/2\pi\approx -0.724$ and
we have $k=1$ from  $\phi'_m=\pi$ and the relation in Eq. (\ref{k}).
Then weak current $I'$ in the connecting loop flows
satisfying current conservation condition between left qubit and connecting loop such that
$E_J\sin 0.181(2\pi)=E_{J1}\sin 0.138(2\pi)+2E'_J\sin 0.724\pi$ for $f'=0$.
%Furthermore, since $\phi'_m=\pi$, we can see that $k=1$ from the relation $\phi'_m=(k-f')\pi$.
When $f'$ approaches $0.5$, the effective Josephson coupling energy $\tilde{E}'_J=E'_J\cos\pi f'$ and thus
the current $I'$ in connecting loop become zero, %-(2\pi/\Phi_0)\sum^2_{i=1}E'_J\sin\phi'_i=2\tilde{E}'_J\sin(\phi'_1+\phi'_2)/2=0$
which means that the coupling between two qubits is switched off.

Now we want to explain the case that two control fluxes are in the same directions and
the case that there is only single dc-SQUID in connecting loop.
If two control fluxes are in the same direction such as
\begin{equation}
f'_L=-f'_R=f',
\end{equation}
the Josephson junction energy
of the connecting loop becomes
\begin{eqnarray}
\label{SwiSame}
U_{\rm conn}(\phi'_m,\phi_m)\!=\!4E'_J[1-\cos\pi f'\cos\phi'_m\cos(2\phi_m-\pi f')].
\nonumber\\
\end{eqnarray}
Similar procedure as in the case of opposite directions of control fluxes
shows that  the same spin states, $|\downarrow\downarrow\rangle$ and $|\uparrow\uparrow\rangle$,
have equal energy such as
\begin{eqnarray}
E_{\downarrow\downarrow}=E_{\uparrow\uparrow}
=E_J\left(4-\frac{E_J}{E_{J1}}\right)+4E'_J\sin^2\pi f'.
\end{eqnarray}
for $\cos\phi_p=E_J/2E_{J1}$.

For different spin states, $|\downarrow\uparrow\rangle$ and $|\uparrow\downarrow\rangle$,
%have different energies as shown in Fig. \ref{sameone} (a).
the energies $E_{\uparrow\downarrow}$ and $E_{\downarrow\uparrow}$ are obtained at two local minima
\begin{eqnarray}
U_{\rm conn}(\phi_m,\phi_p=0)=-4E'_J\cos^2\pi f'\cos2\phi_m ~~~~~~~~~~\nonumber\\
-2E'_J\sin2\pi f'\sin2\phi_m+4E'_J, ~~~~~~~~~~~~
\end{eqnarray}
which can be derived from
Eq. (\ref{SwiSame}).  Since the  states,
$|\downarrow\uparrow\rangle$ and $|\uparrow\downarrow\rangle$,
have different sign for $\phi_{s,-s}$,
the second term produces the energy difference
\begin{eqnarray}
\label{deltaE}
\Delta E=E_{\uparrow\downarrow}-E_{\downarrow\uparrow}
=4E'_J\sin2\pi f'|\sin2\phi_{s,-s}|,
\end{eqnarray}
where $\phi_{s,-s}$ is again one of the values of $\phi_{m}$ for the different spin states.

Figure \ref{ContPot}(b) (lower panel) for $f'=0.25$ shows that,
when two control fluxes are in the same direction,
the energies $E_{\uparrow\downarrow}$ and $E_{\downarrow\uparrow}$
are different while $E_{\downarrow\downarrow}= E_{\uparrow\uparrow}$.
The energy levels of $E_{ss'}$ are plotted in Fig. \ref{TwoDiff}(b).
In this case the effective fluxes $h_L$ and $h_R$ applied to left and right
qubits  in the Hamiltonian of Eq. (\ref{Hcoup}) become different
each other, $h_L \neq h_R$, as $f'$ increases from zero.
For the different current state in Fig. \ref{Coup3JJs}(b),
if the control fluxes $f'_L$ and $f'_R$ threading the dc-SQUID loops
are in the same direction,
the increased current $I'$ in the connecting loop will
flow through the left and right qubit loops.
Thus the qubit states are influenced by  additional effective fluxes,
which will makes the two-qubit operations difficult.
However, if two control fluxes $f'_L$ and $f'_R$ are in opposite directions,
the energies of different spin states remains equal to each other,
$E_{\uparrow\downarrow}=E_{\downarrow\uparrow}$, as shown in
Fig. \ref{TwoDiff}(a).  This means that the additional currents
coming from two dc-SQUID's are cancelled each other and
total additional current induced by the control fluxes
$f'_L$ and $f'_R$ is vanishing in the connecting loop.
As a result, the net effect is just  renormalizing the coupling constant $J$
of the coupled qubit system.

We also calculated energies of coupled qubit states with single dc-SQUID loop whose
boundary conditions become
\begin{eqnarray}
\phi'_1=2\pi (r+f'_{\rm ind})+(\phi_{L1}-\phi_{R1})~~~~~~~~~~~~~\\
-\phi'_1+\phi'_2=2\pi(f'+p),~~~~~~~~~~~~~~~~~~~~~~~~
\end{eqnarray}
instead of those in Eqs. (\ref{loopbc}) and (\ref{switchbc}).
Then we get  the  Josephson junction energies  of the dc-SQUID,
\begin{eqnarray}
\label{SwiSingle}
U_{\rm conn}(\phi_m)=-2{\tilde E}'_J\cos(4\phi_m-\pi f')+2E'_J,
\end{eqnarray}
which gives results similar to those of two dc-SQUID's with fluxes in the same direction  such that
\begin{eqnarray}
E_{\downarrow\downarrow}=E_{\uparrow\uparrow}=
E_J\left(4-\frac{E_J}{E_{J1}}\right)+2E'_J\sin^2\pi f'
\end{eqnarray}
for $\cos\phi_p=E_J/2E_{J1}$ and
\begin{eqnarray}
\Delta E=E_{\uparrow\downarrow}-E_{\downarrow\uparrow}
=2E'_J\sin2\pi f'|\sin4\phi_{s,-s}|
\end{eqnarray}
as shown in Fig. \ref{TwoDiff}(c).
Hence the behaviors of one dc-SQUID in the connecting loop
are qualitatively the same as
those of two dc-SQUID's with fluxes in the same direction.
Therefore  we need two control fluxes threading dc-SQUID's in opposite directions
to cancel the additional currents in the connecting loop
for obtaining the controllable coupling.

In order to obtain the controllable coupling
both the qubit operating flux, $f_{L}$,
and control flux, $f'_{L}$, of the left qubit become in opposite direction to those of the
right qubit, $f_R$ and $f'_R$ as shown in Fig. \ref{Coup3JJs}.
In  real experiments it will be very hard to apply magnetic fluxes of different directions
simultaneously.
We have previously suggested a scalable design for phase-coupled flux qubits \cite{Kim},
where an arbitrary pair of qubits are coupled in a twisted way.
Thus just applying all magnetic fluxes in the same direction makes
automatically the effect of fluxes in opposite directions,
removing the experimental difficulty.

The recent experiment on the phase-coupled flux qubits
without dc-SQUID loop \cite{Ploeg} has shown that the coupled qubit states
are in quantum mechanically superposed regime.
The dc-SQUID loops in the connecting loop of the present tunable coupling scheme
may cause a decoherence effect on the coupled qubit states.
A recent study argued that the  dc-SQUID based oscillator
should be the main source of the decoherence of the flux qubits \cite{Bertet2}.
For the scalable design in Ref. \onlinecite{Kim}, however,
the decoherence from two dc-SQUID's can be reduced.
Since two dc-SQUID's are connected in a twisted way,
the fluctuations from tank circuit or flux lines can be cancelled each other.
%However, since  the more severe decoherence of flux qubits comes from the noises in the microwaves
%for qubit oscillations rather than  static fluxes,
%the applied control fluxes, $f'_{L(R)}$, will not produce severe decoherence.

In realistic implementation of qubit operations, operating  external fluxes are slightly different from
the co-resonance point, $f_L=f_R=0.5$, and
moreover we cannot any more neglect small kinetic inductance and induced fluxes.
Hence we confirmed the results in this study  by numerical calculation
using the exact boundary conditions similar to those in Eqs. (\ref{qubitbc})$-$(\ref{switchbc}),
current-phase relation $I_i=-(2\pi/\Phi_0)E_{Ji}\sin\phi_i$ and current conservation conditions \cite{Kim}.

\section{Summary}

Controllable coupling between two phase-coupled flux qubits can be achieved by using
two dc-SQUID's in the connecting loop with threading fluxes in opposite directions.
We analytically show at co-resonance point ($f_L=f_R=0.5$) that the coupling strength of the phase-coupled flux qubits
can be adjusted by varying the threading fluxes $f'$ from $0$ to $0.5$;
it can be as strong as O$(E'_J)$ and zero in switching-off limit.
%varies from sufficiently strong to zero in switching-off limit
When either two control fluxes are in the same directions or there is only one
dc-SQUID in the connecting loop, the coupled qubits cannot be described by the coupling Hamiltonian.
In slightly different parameter regimes of experimental implementations numerical calculations can be done
to obtain exact results.

\begin{center}
{\bf ACKNOWLEDGMENTS}
\end{center}

This work was supported  by the Ministry of Science and Technology of Korea
(Quantum Information Science).


\begin{thebibliography}{1}
\bibitem{Galindo} Y. Makhlin, G. Sch\"on, and A. Shnirman, Rev. Mod. Phys. {\bf 73}, 357 (2001);
A. Galindo and M. A. Mart{\'i}n-Delgado, {\it ibid.}  {\bf 74}, 347 (2002).
\bibitem{Nakamura} Y. Nakamura, Yu. A. Pashkin, and J. S. Tsai, Nature {\bf 398}, 786 (1999).
\bibitem{Yu} Y. Yu, S. Han, X. Chu, S. Chu, and Z. Wang, Science {\bf 296}, 889 (2002).
\bibitem{Chio} I. Chiorescu, Y. Nakamura, C. J. P. M. Harmans, and J. E. Mooij,
Science {\bf 299}, 1869 (2003).
\bibitem{Pashkin} Yu. A. Pashkin, T. Yamamoto, O. Astafiev, Y. Nakamura, D. V. Averin, and J. S. Tsai,
Nature {\bf 421}, 823 (2003).
\bibitem{Yamamoto} T. Yamamoto, Yu. A. Pashkin, O. Astafiev, Y. Nakamura, and J. S. Tsai,
Nature {\bf 425}, 941 (2003).
\bibitem{Izmalkov} A. Izmalkov, M. Grajcar, E. I{\'l}ichev, Th. Wagner, H.-G. Meyer, A.Yu. Smirnov, M. H. S. Amin, Alec Maassen van den Brink, and A.M. Zagoskin,
Phys. Rev. Lett. {\bf 93}, 037003 (2004).
\bibitem{Grajcar} M. Grajcar, A. Izmalkov, S. H. W. van der Ploeg, S. Linzen, E. I{\'l}ichev, Th. Wagner, U. Hubner, H.-G. Meyer,
Alec Maassen van den Brink, S. Uchaikin, and A. M. Zagoskin,
Phys. Rev. B {\bf 72}, 020503(R) (2005).
\bibitem{Majer} J. B. Majer, F. G. Paauw, A. C. J. ter Haar, C. J. P. M. Harmans, and J. E. Mooij,
Phys. Rev. Lett. {\bf 94}, 090501 (2005).
\bibitem{Berkley} A. J. Berkley {\it et al.}, %H. Xu, R. C. Ramos, M. A. Gubrud, F. W. Strauch, P. R. Johnson,J. R. Anderson, A. J. Dragt, C. J. Lobb, and F. C. Wellstood,
Science {\bf 300}, 1548 (2003).
%\bibitem{You} J. Q. You, Y. Nakamura, and F. Nori, Phys. Rev. B {\bf 71}, 024532 (2005).
\bibitem{Averin} D. V. Averin and C. Bruder, Phys. Rev. Lett. {\bf 91}, 057003 (2003);
J. Q. You, J. S. Tsai, and F. Nori, Phys. Rev. B {\bf 68}, 024510 (2003).
\bibitem{Blais} A. Blais, Alec Maassen van den Brink, and A. M. Zagoskin, Phys. Rev. Lett. {\bf 90}, 127901 (2003).
\bibitem{Plourde} B. L. T. Plourde {\it et al.}, %J. Zhang, K. B. Whaley, F. K. Wilhelm, T. L. Robertson, T. Hime, S. Linzen, P. A. Reichardt, C.-E. Wu, and J. Clarke,
Phys. Rev. B {\bf 70}, 140501(R) (2004).
\bibitem{Bertet} P. Bertet, C. J. P. M. Harmans, and J. E. Mooij, Phys. Rev. B {\bf 73}, 064512 (2006).
\bibitem{Liu} Y.-x. Liu, L. F. Wei, J. S. Tsai, and F. Nori, Phys. Rev. Lett. {\bf 96}, 067003 (2006).
\bibitem{Niskanen} A. O. Niskanen, Y. Nakamura, and J. S. Tsai, Phys. Rev. B {\bf 73}, 094506 (2006).

\bibitem{Kim} M. D. Kim and J. Hong, Phys. Rev. B {\bf 70}, 184525 (2004).

\bibitem{Ploeg} S. H. W. van der Ploeg, A. Izmalkov,  Alec Maassen van den Brink,
U. H{\" u}bner, M. Grajcar, E. I{\'l}ichev, H.-G. Meyer, and A.M. Zagoskin, cond-mat/0605588.

\bibitem{Grajcar2} M. Grajcar, Y.-x. Liu, F. Nori, and A. M. Zagoskin, cond-mat/0605484.

\bibitem{Grajcar3} M. Grajcar {\it et al.}, %A. Izmalkov,1 S. H. W. van der Ploeg,1,4 S. Linzen,1
%T. Plecenik,1,2 Th. Wagner,1 U. Hubner,1 E. Il'ichev,1 H.-G. Meyer,1 A. Yu. Smirnov,5
%Peter J. Love,5 Alec Maassen van den Brink,5 M. H. S. Amin,5 S. Uchaikin,5 and A. M. Zagoskin
Phys. Rev. Lett. 96, 047006 (2006).

\bibitem{Brink} A. Maassen van den Brink, cond-mat/0605398.

\bibitem{Mooij} J. E. Mooij {\it et al.}, %T. P. Orlando, L. Levitov, Lin Tian, Caspar H. van der Wal, and Seth Lloyd,
Science {\bf 285}, 1036 (1999);
Caspar H. van der Wal {\it et al.}, %A. C. J. ter Haar, F. K. Wilhelm, R. N. Schouten, C. J. P. M. Harmans, T. P. Orlando, Seth Lloyd, and J. E. Mooij,
Science {\bf 290}, 773 (2000).
\bibitem{Orlando}  T. P. Orlando {\it et al.}, %J. E. Mooij, Lin Tian, Caspar H. van der Wal, L. S. Levitov, Seth Lloyd, and J. J. Mazo,
Phys. Rev. B {\bf 60}, 15398 (1999).
\bibitem{Kim1} M. D. Kim, D. Shin, and J. Hong, Phys. Rev. B {\bf 68}, 134513 (2003).

\bibitem{Bertet2} P. Bertet, I. Chiorescu, G. Burkard, K. Semba,
C. J. P. M. Harmans, D. P. DiVincenzo, and J. E. Mooij, Phys. Rev. Lett. {\bf 95}, 257002 (2005).
%
%\bibitem{Astafiev} O. Astafiev, Yu. A. Pashkin, Y. Nakamura, T. Yamamoto, and J. S. Tsai,
%Phys. Rev. Lett. {\bf 96}, 137001 (2006).
%\bibitem{Yoshihara} F. Yoshihara, K. Harrabi, A. O. Niskanen, Y. Nakamura, and J. S. Tsai,
%cond-mat/0606481.
\end{thebibliography}
\end{document}